\documentstyle[epsf]{article}

\begin{document}

\begin{center}

{\Large\bf Estimating long range dependence:\\ 
finite sample properties and confidence intervals}

\vspace*{.6cm}

{\large Rafa{\l} Weron}

\vspace*{.2cm}

{\small
Hugo Steinhaus Center for Stochastic Methods,\\
Wroc{\l}aw University of Technology, 50-370 Wroc{\l}aw, Poland\\
E-mail: rweron@im.pwr.wroc.pl
}

\end{center}

\vspace*{.5cm}

\noindent
{\bf Abstract:}
A major issue in financial economics is the behavior of asset returns over long horizons.
Various estimators of long range dependence have been proposed. Even though some have 
known asymptotic properties, it is important to test their accuracy by using simulated 
series of different lengths. 
We test R/S analysis, Detrended Fluctuation Analysis and periodogram regression methods
on samples drawn from Gaussian white noise. The DFA statistics turns out to be the unanimous 
winner. Unfortunately, no asymptotic distribution theory has been derived for this 
statistics so far. We were able, however, to construct empirical (i.e. approximate) 
confidence intervals for all three methods. The obtained values differ largely from 
heuristic values proposed by some authors for the R/S statistics and are very close 
to asymptotic values for the periodogram regression method.

\vspace*{.3cm}


\vspace*{.5cm}


\section{Introduction}

Time series with long-range dependence are widespread in nature and for many years have been 
extensively studied in hydrology and geophysics \cite{hurst51,mw68,mw69,mandelbrot82}. 
More recently, "long memory" or "$1/f$ noise" (as it is often called) has been observed 
in DNA sequences \cite{peng_etal92,peng_etal94}, cardiac dynamics 
\cite{peng_etal93,makikallio_etal99}, internet traffic \cite{wte96}, meteorology 
\cite{k-b_etal98,mrt00}, geology \cite{mt99} and even ethology \cite{ah00}. 

In economics and finance, long-range dependence also has a long history (for a review see 
Refs. \cite{bk96,mandelbrot97}) and still is a hot topic of active research 
\cite{lux96,ww98,ls98,wtt99,g-c00,lv00,wp00,clk00,mandelbrot01}. 
Historical records of economic and financial data typically exhibit nonperiodic cyclical 
patterns that are indicative of the presence of significant long memory. However, the 
statistical investigations that have been performed to test long-range dependence have 
often become a source of major controversies, especially in the case of stock returns. 
The reason for this are the implications that the presence of long memory has on many 
of the paradigms used in modern financial economics \cite{wtt99,lo91}.

Various estimators of long range dependence have been proposed. Even though some have 
known asymptotic properties, it is important to test their accuracy by using simulated 
series of different lengths. Such a study was presented in Refs. \cite{ttw95,tt98} using 
"ideal" models that display long-range dependence, i.e fractional Brownian noise (fBn) 
and Fractional ARIMA(0,d,0). However, the authors tested estimators on rather long 
time series (10000 elements), whereas in practice we often have to perform analysis of 
much shorter data sets. For example, Lux \cite{lux96} used daily time series comprising 
1949 DAX returns, Lobato and Savin \cite{ls98} originally performed tests on 8178 S\&P500 
daily returns, but later due to the non-stationarity of the process (which influenced 
the estimates) divided the sample into smaller subsamples, Grau-Carles \cite{g-c00} 
analyzed various stock index data including 4125 Nikkei and 1555 FTSE daily returns,
Weron and Przyby{\l}owicz \cite{wp00} estimated the Hurst exponent using only 670 daily 
returns of the spot electricity price in California.
Moreover, Taqqu et al. \cite{ttw95} based their statistical conclusions on only 50 
simulated trajectories of fBn or FARIMA. This may be enough for the estimation of the mean 
or median, but certainly not enough for the estimation of very high or low quantiles, which 
can be used to construct empirical confidence intervals \cite{stephens74}.

In this paper we analyze rescaled range analysis, Detrended Fluctuation Analysis and 
periodogram regression methods on samples drawn from Gaussian white noise. 
In Section 2 we precisely describe all three methods and later, in Section 3, present 
a comparison based on Monte Carlo simulations. The DFA statistics turns out to be the 
unanimous winner. Unfortunately, no asymptotic distribution theory has been derived 
for this statistics so far. We were able, however, to construct empirical (i.e. 
approximate) confidence intervals for all three methods. These results are presented 
in Section 4. The obtained values differ largely from heuristic values proposed by 
some authors for the R/S statistics and are very close to asymptotic values for 
the periodogram regression method. In Section 5 we apply the results of Section 4 
to a number of financial data illustrating their usefulness.

\section{Methods for estimating $H$}

\subsection{R/S analysis}

We begun our investigation with one of the oldest and best-known methods, the so-called
R/S analysis. This method, proposed by Mandelbrot and Wallis \cite{mw69b} and based on
previous hydrological analysis of Hurst \cite{hurst51}, allows the calculation of the 
self-similarity parameter $H$, which measures the intensity of long range dependence 
in a time series. 

The analysis begins with dividing a time series (of returns) of length $L$ into $d$ 
subseries of length $n$. Next for each subseries $m=1,...,d$: 
1$^{\circ}$ find the mean ($E_m$) and standard deviation ($S_m$); 
2$^{\circ}$ normalize the data ($Z_{i,m}$) by subtracting the sample mean 
$X_{i,m}=Z_{i,m}-E_m$ for $i=1,...,n$;
3$^{\circ}$ create a cumulative time series $Y_{i,m}=\sum_{j=1}^i X_{j,m}$ for $i=1,...,n$;
4$^{\circ}$ find the range $R_m = \max \{Y_{1,m},...,Y_{n,m}\} - \min \{Y_{1,m},...,Y_{n,m}\}$;
and 5$^{\circ}$ rescale the range $R_m/S_m$.
Finally, calculate the mean value of the rescaled range for all subseries of length $n$ 
$$
(R/S)_n  = \frac{1}{d}\sum_{m=1}^d R_m/S_m.
$$ 

It can be shown \cite{mandelbrot75} that the R/S statistics asymptotically follows the relation
$$
(R/S)_n \sim cn^H.
$$
Thus the value of $H$ can be obtained by running a simple linear regression over a sample of
increasing time horizons 
$$
\log (R/S)_n = \log c + H \log n.
$$
Equivalently, we can plot the $(R/S)_n$ statistics against $n$ on a double-logarithmic 
paper, see Fig. 1.
If the returns process is white noise then the plot is roughly a straight line with slope 0.5. 
If the process is persistent then the slope is greater than 0.5; if it is anti-persistent then 
the slope is less than 0.5. The "significance" level is usually chosen to be one over the square 
root of sample length, i.e. the standard deviation of a Gaussian white noise \cite{peters94}.

\begin{figure}[tbp]
\centerline{\epsfxsize=10cm \epsfbox{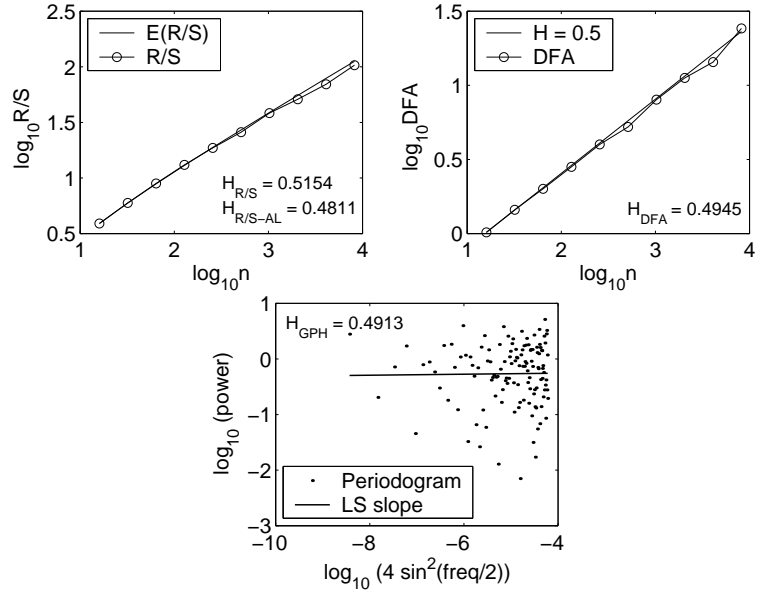}}
\caption{Estimation of the Hurst exponent for a white noise ($H=0.5$) sequence 
of $2^{14}=16384$ observations via the Hurst R/S analysis ({\it top left panel}),
the Detrended Fluctuation Analysis ({\it top right panel})
and the periodogram Geweke-Porter-Hudak method ({\it bottom panel}).
}
\end{figure}

However, it should be noted that for small $n$ there is a significant deviation from 
the 0.5 slope. For this reason the theoretical (i.e. for white noise) values of the R/S 
statistics are usually approximated by 
\begin{equation}\label{eqn-ERS}
{\bf E}(R/S)_n = \left\{
  \begin{array}{lll}
  \frac{n-\frac{1}{2}}{n} \frac{\Gamma(\frac{n-1}{2})}{\sqrt{\pi} \Gamma(\frac{n}{2})}
    \sum\limits_{i=1}^{n-1} \sqrt{\frac{n-i}{i}} & \mbox{for} & n\le 340, \\
  \frac{n-\frac{1}{2}}{n} \frac{1}{\sqrt{n\frac{\pi}{2}}}
    \sum\limits_{i=1}^{n-1} \sqrt{\frac{n-i}{i}} & \mbox{for} & n>340, \\
  \end{array} \right.
\end{equation}
where $\Gamma$ is the Euler gamma function. This formula is a slight modification of 
the formula given by Anis and Lloyd \cite{al76}; the $(n-\frac12)/n$ term was added by 
Peters \cite{peters94} to improve the performance for very small $n$. 

Formula (\ref{eqn-ERS}) was used as a benchmark in all empirical studies of the R/S 
statistics presented in this paper, i.e. the Hurst exponent $H$ was calculated as $0.5$ 
plus the slope of $(R/S)_n - {\bf E}(R/S)_n$. The resulting statistics was denoted by R/S-AL.

A major drawback of the R/S analysis is the fact that no asymptotic distribution 
theory has been derived for the Hurst parameter $H$. The only known results are for the 
rescaled (but not by standard deviation) range $R_m$ itself \cite{lo91}.

\subsection{Detrended Fluctuation Analysis}

The second method we used to measure long range dependence was the Detrended Fluctuation 
Analysis (DFA) proposed by Peng et al. \cite{peng_etal94}. The method can be summarized 
as follows. Divide a time series (of returns) of length $L$ into $d$ subseries of length $n$. 
Next for each subseries $m=1,...,d$: 
1$^{\circ}$ create a cumulative time series $Y_{i,m}=\sum_{j=1}^i X_{j,m}$ for $i=1,...,n$;
2$^{\circ}$ fit a least squares line $\tilde{Y}_m(x)=a_m x + b_m$ to $\{Y_{1,m},...,Y_{n,m}\}$;
and 3$^{\circ}$ calculate the root mean square fluctuation (i.e. standard deviation) 
of the integrated and detrended time series 
$$
F(m) = \sqrt{\frac1n \sum_{i=1}^n (Y_{i,m} - a_m i - b_m)^2}.
$$
Finally, calculate the mean value of the root mean square fluctuation for all subseries of 
length $n$ 
$$
\bar{F}(n)=\frac1d \sum_{m=1}^d F(m).
$$ 
Like in the case of R/S analysis, a linear relationship on a double-logarithmic paper 
of $\bar{F}(n)$ against the interval size $n$ indicates the presence of a power-law scaling of 
the form $cn^H$ \cite{peng_etal94,ttw95}. If the returns process is white noise then the 
slope is roughly 0.5. If the process is persistent then the slope is greater than 0.5; 
if it is anti-persistent then the slope is less than 0.5. 

Unfortunately, no asymptotic distribution theory has been derived for the DFA statistics so far. 
Hence, no explicit hypothesis testing can be performed and the significance relies on subjective 
assessment.

\subsection{Periodogram regression}

The third method is a semi-parametric procedure to obtain an estimate of the fractional 
differencing parameter $d$. This technique, proposed by Geweke and Porter-Hudak \cite{gp-h83} 
(GPH), is based on observations of the slope of the spectral density function of a fractionally 
integrated series around the angular frequency $\omega = 0$. Since they showed that the spectral 
density function of a general fractionally integrated model (eg. FARIMA) with differencing 
parameter $d$ is identical to that of a fractional Gaussian noise with Hurst exponent $H=d+0.5$, 
the GPH method can be used to estimate $H$.

The estimation procedure begins with calculating the periodogram, which is a sample analogue 
of the spectral density. For a vector of observations $\{x_1,...,x_L\}$ the periodogram is 
defined as
$$
I_L(\omega_k)=\frac{1}{L} \left|\sum_{t=1}^{L} x_t e^{-2\pi i (t-1) \omega_k} \right|^2,
$$
where $\omega_k = k/L$, $k=1,...,[L/2]$ and $[x]$ denotes the largest integer less then or 
equal to $x$. Observe that $I_L$ is the squared absolute value of the Fourier transform
and if the observations vector is of appropriate length (even or a power of 2) 
then we can use fast algorithms to calculate the Fourier transform. 

The next and final step is to run a simple linear regression 
\begin{equation}\label{gph-reg}
\log\{I_L(\omega_k)\} 
= a - \hat{d} \log\left\{4 \sin^2 (\omega_k/2) \right\} + \epsilon_k,
\end{equation}
at low Fourier frequencies $\omega_k$, $k=1,...,K \le [L/2]$. The least squares estimate 
of the slope yields the differencing parameter $d$ through the relation $d = \hat{d}$, 
hence $H = \hat{d}+0.5$.
A major issue on the application of this method is the choice of $K$. Geweke and Porter-Hudak 
\cite{gp-h83}, as well as a number of other authors, recommend choosing $K$ such that 
$K=[L^{0.5}]$, however, other values (eg. $K=[L^{0.45}]$, $[L^{0.2}]\le K\le [L^{0.5}]$) have 
also been suggested.

Periodogram regression is the only of the presented methods, which has known asymptotic 
properties. Inference is based on the asymptotic distribution of the estimate $\hat{d}$ 
of $d$, which is normally distributed with mean $d$ and variance 
\begin{equation}\label{gph_var}
\frac{\pi^2}{6 \sum_{k=1}^K (x_t - \bar{x})^2},
\end{equation}
where $x_t = \log\{4 \sin^2 (\omega_k/2)\}$ is the regressor in eq. (\ref{gph-reg}).

\section{Comparison of estimators}

In order to test the presented estimation methods we performed Monte Carlo simulations.
We generated samples of Gaussian white noise sequences (independent and $N(0,1)$ distributed) 
of length $L=2^N$, where $N=8,9,...,16$, i.e. $L=256, 512, ..., 65536$. For each $L$ ten 
thousand trajectories were produced. Next, we applied all three estimation procedures --
Anis-Lloyd corrected R/S statistics, Detrended Fluctuation Analysis and periodogram 
regression -- to the data series and compared the results. 

\begin{figure}[tbp]
\centerline{\epsfxsize=10cm \epsfbox{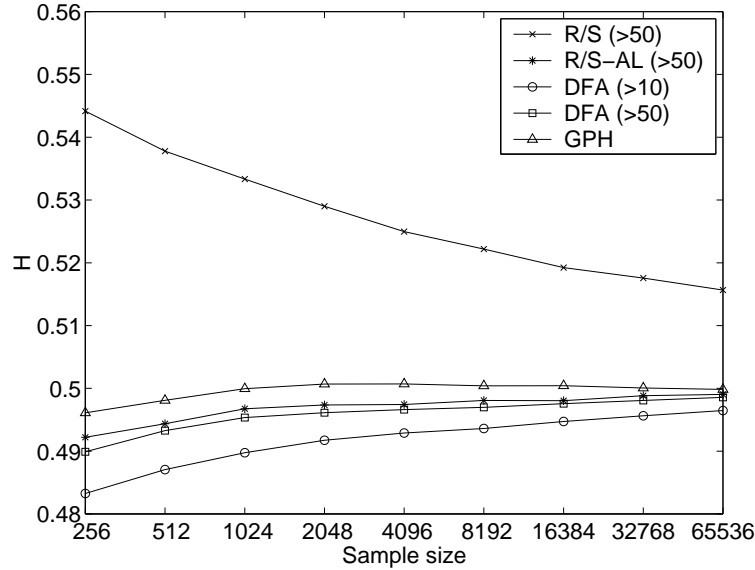}}
\caption{Mean values (over 10000 samples) of the estimated Hurst exponents for Gaussian 
white noise sequences of length $L = 256, 512, ..., 65536$. The methods used are
the classical R/S analysis (R/S) for subintervals of length $n>50$, 
the Anis-Lloyd corrected R/S statistics (R/S-AL) for $n>50$, 
the Detrended Fluctuation Analysis for $n>10$ and $n>50$,
and the periodogram Geweke-Porter-Hudak method for cutoff $K=[L^{0.5}]$.
}
\end{figure}

In Figure 2 we plotted the mean values (over all 10000 samples) of the estimated Hurst 
exponents. For illustrative purposes we also included the results of the classical R/S 
statistics (R/S), i.e. without the Anis-Lloyd correction. As can be seen in Fig. 2, on 
average, this method overestimates the true Hurst exponent to a great extent. 
On the other hand, all other methods have a slight negative bias. 

We analyzed the R/S and DFA methods for two cutoffs of the subinterval length: $n>10$ 
and $n>50$. Despite the corrections to the original rescaled range statistics, R/S-AL 
still possesses a large variance if very small $n$ are included in the calculations. 
Thus we decided to use only subintervals of length $n>50$ (to be more precise: 
$n=64,128,256,...$, since the length of our samples was a power of 2). 
On the other hand, the DFA statistics behaves nicely even for small subintervals. 
So, we calculated $H_{DFA}$ for $n>10$ (more precisely: $n=16,32,64,...$) and -- to be 
consistent with the results for the rescaled range analysis -- separately for $n>50$.

As we have mentioned in the previous Section, results of the Geweke-Porter-Hudak method 
depend on the choice of the cutoff value $K$, which determines how many of the low Fourier
frequencies $\omega_k$ are taken into account. In our simulations we decided to use 
the standard value, i.e. $K=[L^{0.5}]$, because lower powers of $L$ introduce larger 
estimation errors, while larger powers force us to move away from the region for which 
the theoretical results hold.

The large number of simulated trajectories allowed us to compare the methods. For a given 
method we obtained 10000 estimated values of $H$, called $\{H_i, i=1,2,...,10000\}$.
Apart from calculating their mean (see Fig. 2), we computed their standard deviation
and mean absolute error, i.e.
$$
\mbox{MAE} = \frac{1}{10000} \sum_{i=1}^{10000} |H_i - 0.5|,
$$
which provides some information on the bias. The results are presented in Table 1.
To have a complete picture, in Fig. 3 we also plotted the 2.5\% and 97.5\% sample 
quantiles for all methods. A $p\%$ sample quantile, denoted by $\hat{x}_p$, is such 
a value that $p\%$ of the sample observations are less than $\hat{x}_p$. Equivalently 
we can say that the empirical distribution $F_e$ evaluated at $\hat{x}_p$ equals $p\%$,
i.e. $F_e(\hat{x}_p)=p\%$.

\begin{figure}[tbp]
\centerline{\epsfxsize=10cm \epsfbox{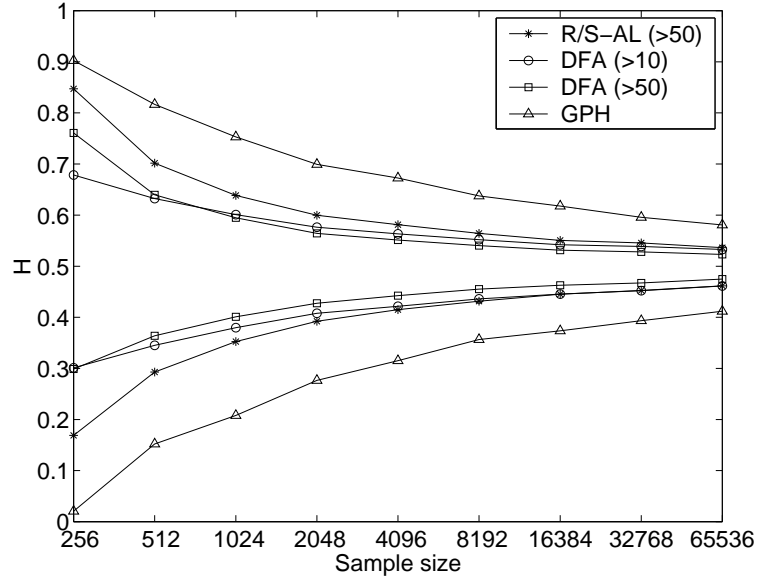}}
\caption{2.5\% and 97.5\% sample quantiles (over 10000 samples) of the estimated Hurst 
exponents for Gaussian white noise sequences of length $L = 256, 512, ..., 65536$. 
The methods used are the Anis-Lloyd corrected R/S statistics (R/S-AL) for $n>50$, 
the Detrended Fluctuation Analysis for $n>10$ and $n>50$,
and the periodogram Geweke-Porter-Hudak method for cutoff $K=[L^{0.5}]$.
}
\end{figure}

\begin{table}[tbp]
\caption{Estimation results for $H=0.5$ using 10000 independent realizations of different length.
The best results in each category are in bold.
}
\begin{center}
\begin{tabular}{ccccc}
\hline
Sample& \multicolumn{4}{c}{Method} \\
size  & R/S-AL ($n>50$) & DFA ($n>10$) & DFA ($n>50$) & GPH\\
\hline
      & \multicolumn{4}{c}{Standard deviation} \\
\hline
  256 &    0.1739  &  {\bf 0.0972}  &  0.1222  &  0.2205\\
  512 &    0.1043  &  0.0724  &  {\bf 0.0716}  &  0.1705\\
 1024 &    0.0739  &  0.0559  &  {\bf 0.0497}  &  0.1401\\
 2048 &    0.0535  &  0.0436  &  {\bf 0.0355}  &  0.1085\\
 4096 &    0.0422  &  0.0359  &  {\bf 0.0278}  &  0.0898\\
 8192 &    0.0332  &  0.0293  &  {\bf 0.0217}  &  0.0727\\
16384 &    0.0277  &  0.0254  &  {\bf 0.0181}  &  0.0622\\
32768 &    0.0237  &  0.0223  &  {\bf 0.0154}  &  0.0511\\
65536 &    0.0193  &  0.0184  &  {\bf 0.0125}  &  0.0424\\
\hline
      & \multicolumn{4}{c}{Mean absolute error} \\
\hline
  256 &    0.1396  &  {\bf 0.0800}  &  0.1003  &  0.1733\\
  512 &    0.0838  &  0.0589  &  {\bf 0.0582}  &  0.1357\\
 1024 &    0.0594  &  0.0455  &  {\bf 0.0401}  &  0.1108\\
 2048 &    0.0428  &  0.0355  &  {\bf 0.0287}  &  0.0861\\
 4096 &    0.0337  &  0.0291  &  {\bf 0.0224}  &  0.0711\\
 8192 &    0.0266  &  0.0240  &  {\bf 0.0176}  &  0.0580\\
16384 &    0.0224  &  0.0209  &  {\bf 0.0147}  &  0.0494\\
32768 &    0.0189  &  0.0181  &  {\bf 0.0124}  &  0.0402\\
65536 &    0.0155  &  0.0150  &  {\bf 0.0101}  &  0.0335\\
\hline
\end{tabular}
\end{center}
\end{table}

In all three tests the clear winner for $L>500$ is the DFA statistics with $n>50$ as
it gives the estimated values closest to the initial Hurst exponent ($H=0.5$).
The reason it performs worse then the DFA statistics with $n>10$ for the smallest 
tested samples is probably the fact that $L=256$ has only three divisors greater 
than 50: $n=64,128,256$ and that regression based on only three points can yield 
large errors. 

Unfortunately, no asymptotic distribution theory has been derived for the DFA statistics 
so far. However, using Monte Carlo simulations we were able to construct empirical 
(i.e. approximate) confidence intervals for all three analyzed methods.

\section{Construction of confidence intervals}

The estimation of the Hurst exponent $H$ alone is not enough. We also need a measure 
of the significance of the results. Traditionally, the statistical approach is to test
the null hypothesis of no or weak dependence versus the alternative of strong dependence
or long memory at some given significance level. However, to construct a test the asymptotic
distribution of the test statistics must be known. Of the three analyzed methods only 
the spectral one has well know asymptotic properties. In this Section we will thus construct 
empirical confidence intervals.

The procedure is quite simple, although burdensome, and consists of the following: 
1$^{\circ}$ for a set of sample lengths (in our case: $L=256,512,...,65536$) generate 
a large number (here: 10000) of realizations of an independent or a weakly dependent 
time series (here: Gaussian white noise);
2$^{\circ}$ compute the lower (0.5\%, 2.5\%, 5\%) and upper (95\%, 97.5\%, 99.5\%)
sample quantiles for all sample lengths;
and 3$^{\circ}$ plot the sample quantiles vs. sample size and fit them with some functions.
These functions can be later used to construct confidence intervals. The 5\% and 95\%
quantiles designate the 90\% (two-sided) confidence interval, 2.5\% and 97.5\%
quantiles -- the 95\% confidence interval and 0.5\% and 99.5\% quantiles 
-- the 99\% confidence interval.

Results of the above procedure for the Anis-Lloyd corrected R/S statistics are presented 
in Table 2 and Figs. 4-5. To find the 95\% confidence interval we plotted 2.5\% and 97.5\%
sample quantiles vs. sample size. The only satisfactory results were obtained for 
$\log(\hat{x}_{97.5} - 0.5)$ and $\log(0.5 - \hat{x}_{2.5})$ vs. $\log(\log N)$, where
$N = \log_2 L$, see Fig. 4. For sample size $L=256$ the 97.5\% quantile introduced large 
estimation errors, probably due to the fact that $L=256$ has only three divisors greater 
than 50: $n=64,128,256$ and that regression based on only three points can yield 
large errors. Thus we decided to use 97.5\% quantiles (in fact 95\% and 99.5\% 
quantiles as well) coming only from samples of at least 512 observations. The obtained 
fit was very good, the $R^2$ statistics for the lower quantile was 0.9987 and for the 
upper 0.9972. The complete formulas, also for 90\% and 99\% confidence intervals are 
given in Table 2. In Figure 5 we plotted the mean value of $H$ and the 95\% confidence 
interval vs. sample size. For comparison we also added the heuristic "significance level" 
(one over the square root of sample length) given in Peters \cite{peters94}. It is clearly 
seen that this "significance level" rejects the null hypothesis too often compared to the 
95\% confidence interval (in fact too often even compared to the 90\% confidence interval). 

\begin{table}[htbp]
\caption{Empirical confidence intervals for the Anis-Lloyd corrected R/S statistics
and sample length $L=2^N$.
}
\begin{center}
\begin{tabular}{ccc}
\hline
      & \multicolumn{2}{c}{Confidence intervals for $n>50$} \\
Level & Lower bound & Upper bound \\
\hline
90\%  & $0.5 - \exp(-7.35 \cdot \log(\log N) + 4.06)$ 
      & $\exp(-7.07 \cdot \log(\log N) + 3.75) + 0.5$  \\
95\%  & $0.5 - \exp(-7.33 \cdot \log(\log N) + 4.21)$ 
      & $\exp(-7.20 \cdot \log(\log N) + 4.04) + 0.5$  \\
99\%  & $0.5 - \exp(-7.19 \cdot \log(\log N) + 4.34)$ 
      & $\exp(-7.51 \cdot \log(\log N) + 4.58) + 0.5$  \\
\hline
\end{tabular}
\end{center}
\end{table}

\begin{figure}[p]
\centerline{\epsfxsize=12cm \epsfbox{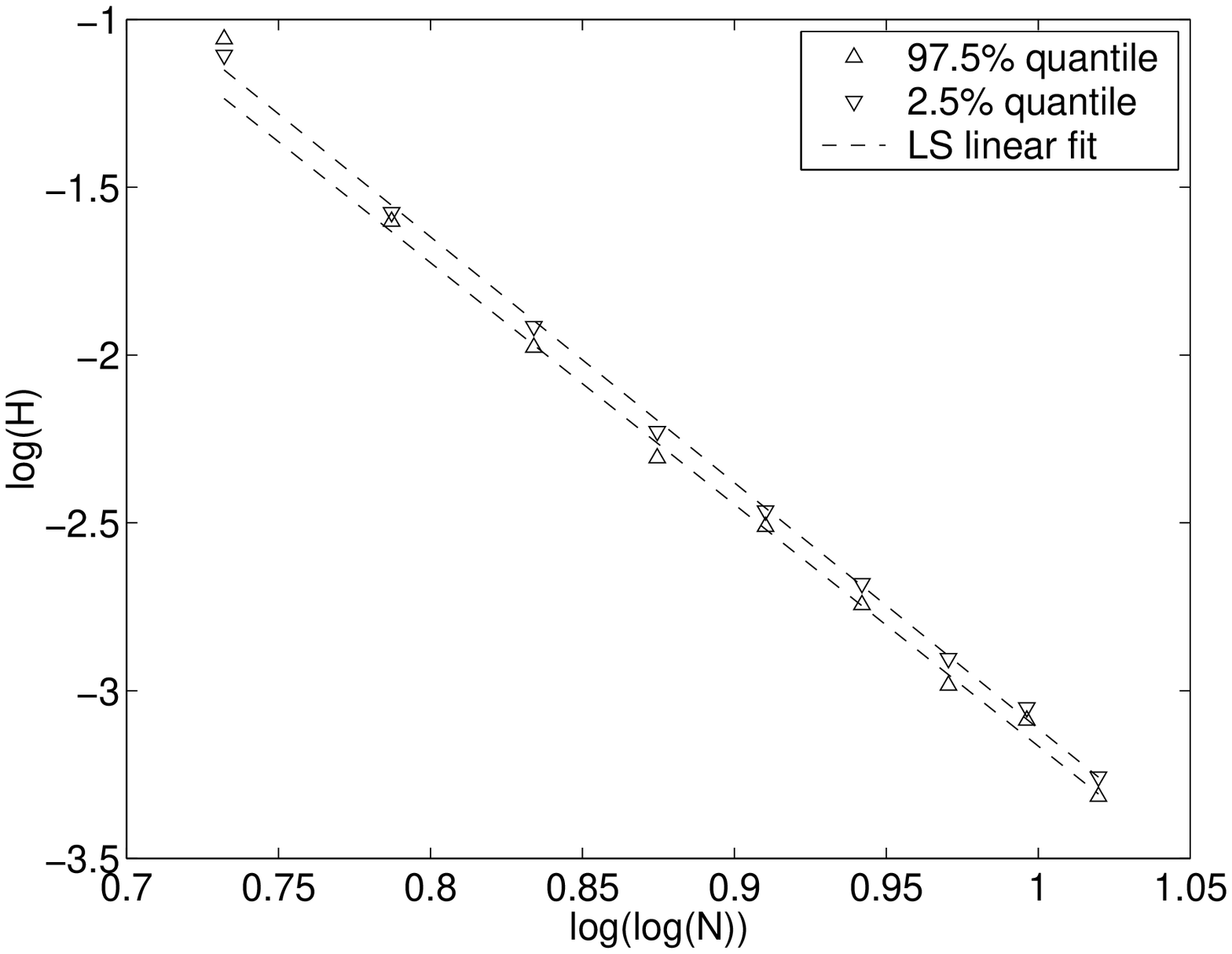}}
\caption{Fitting R/S-AL quantiles ($n>50$): a plot of $\log(\hat{x}_{97.5} - 0.5)$ and 
$\log(0.5 - \hat{x}_{2.5})$ vs. $\log(\log N)$, where $N = \log_2 L$.
For the smallest analyzed samples ($L=256$) the 97.5\% quantile introduced large 
estimation errors and was not used in the analysis.
}

\centerline{\epsfxsize=12cm \epsfbox{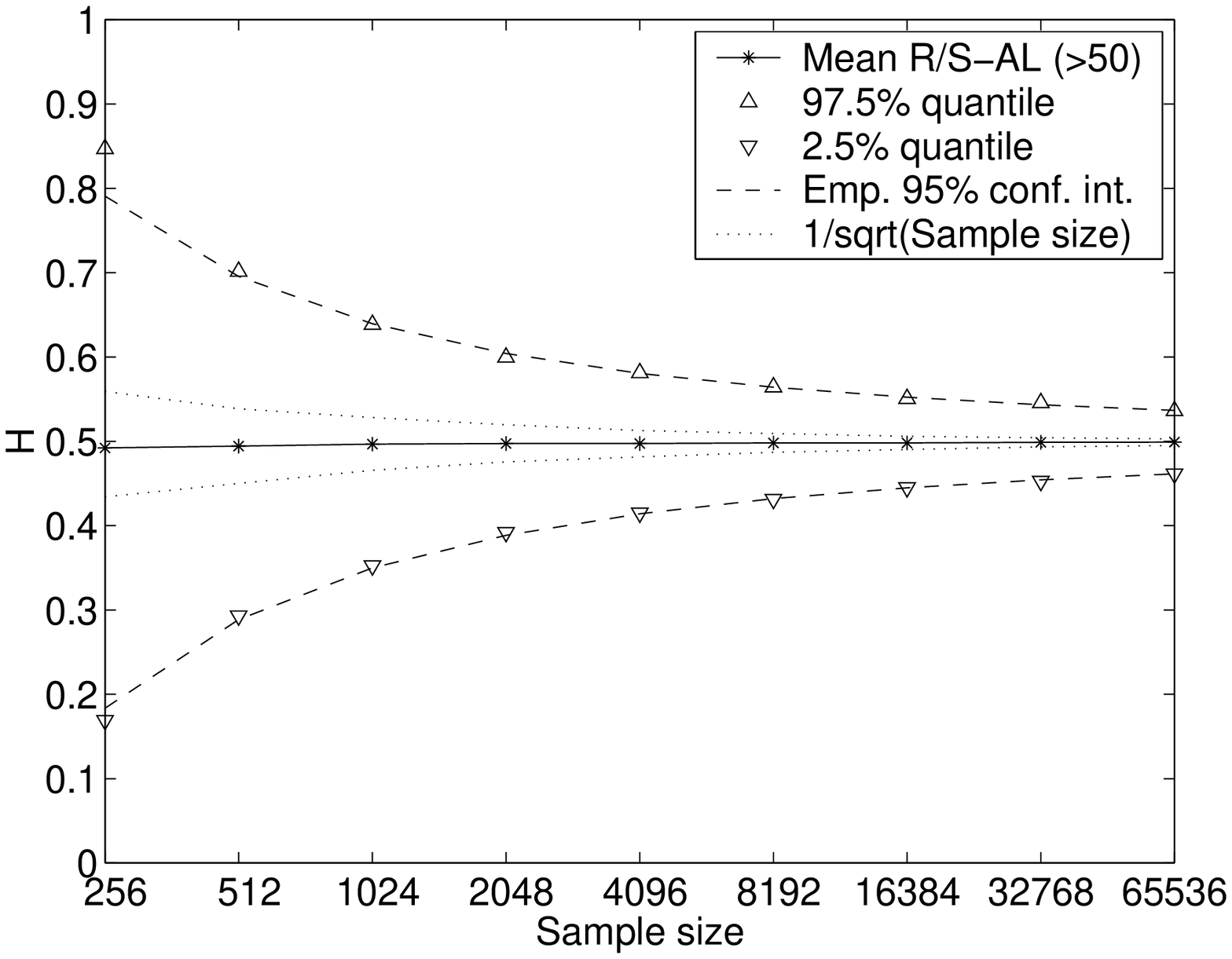}}
\caption{Mean value of the Anis-Lloyd corrected R/S statistics, 95\% empirical confidence 
intervals and $1/\sqrt{L}$ intervals for samples of size $L=256, 512, ..., 65536$.
}
\end{figure}

Results for the DFA statistics are presented in Table 3 and Figs. 6-9. Again in order 
to find the 95\% confidence interval we plotted 2.5\% and 97.5\% sample quantiles vs. 
sample size. The only satisfactory results were obtained for $\log(\hat{x}_{97.5} - 0.5)$ 
and $\log(0.5 - \hat{x}_{2.5})$ against $\log N$, where $N = \log_2 L$, see Figs. 6 and 8. 
Like in the case of the R/S-AL statistics, for sample size $L=256$ the 97.5\% quantile 
introduced large estimation errors when the subinterval size was restricted to $n>50$,
see Fig. 8. Thus we decided to use 97.5\% quantiles (95\% and 99.5\% quantiles as well) 
coming only from samples of at least 512 observations when the higher ($n>50$) cutoff
was applied. The obtained fit was very good. In the $n>10$ case the $R^2$ statistics for 
the lower quantile was 0.9985 and for the upper 0.9972. In the $n>50$ case the $R^2$ 
statistics for the lower quantile was 0.9973 and for the upper 0.9918. The complete 
formulas, also for 90\% and 99\% confidence intervals are given in Table 3. In Figures 7 
and 9 we plotted the mean value of $H$ and the 95\% confidence interval vs. sample size.

\begin{table}[htbp]
\caption{Empirical confidence intervals for the DFA statistics and sample length $L=2^N$.
}
\begin{center}
\begin{tabular}{ccc}
\hline
      & \multicolumn{2}{c}{Confidence intervals for $n>10$} \\
Level & Lower bound & Upper bound \\
\hline
90\%  & $0.5 - \exp(-2.33 \cdot \log N + 3.09)$ 
      & $\exp(-2.44 \cdot \log N + 3.13) + 0.5$  \\
95\%  & $0.5 - \exp(-2.33 \cdot \log N + 3.25)$ 
      & $\exp(-2.46 \cdot \log N + 3.38) + 0.5$  \\
99\%  & $0.5 - \exp(-2.20 \cdot \log N + 3.18)$ 
      & $\exp(-2.45 \cdot \log N + 3.62) + 0.5$  \\
\hline
      & \multicolumn{2}{c}{Confidence intervals for $n>50$} \\
Level & Lower bound & Upper bound \\
\hline
90\%  & $0.5 - \exp(-2.99 \cdot \log N + 4.45)$ 
      & $\exp(-3.09 \cdot \log N + 4.57) + 0.5$  \\
95\%  & $0.5 - \exp(-2.93 \cdot \log N + 4.45)$ 
      & $\exp(-3.10 \cdot \log N + 4.77) + 0.5$  \\
99\%  & $0.5 - \exp(-2.67 \cdot \log N + 4.06)$ 
      & $\exp(-3.19 \cdot \log N + 5.28) + 0.5$  \\
\hline
\end{tabular}
\end{center}
\end{table}

\begin{figure}[p]
\centerline{\epsfxsize=12cm \epsfbox{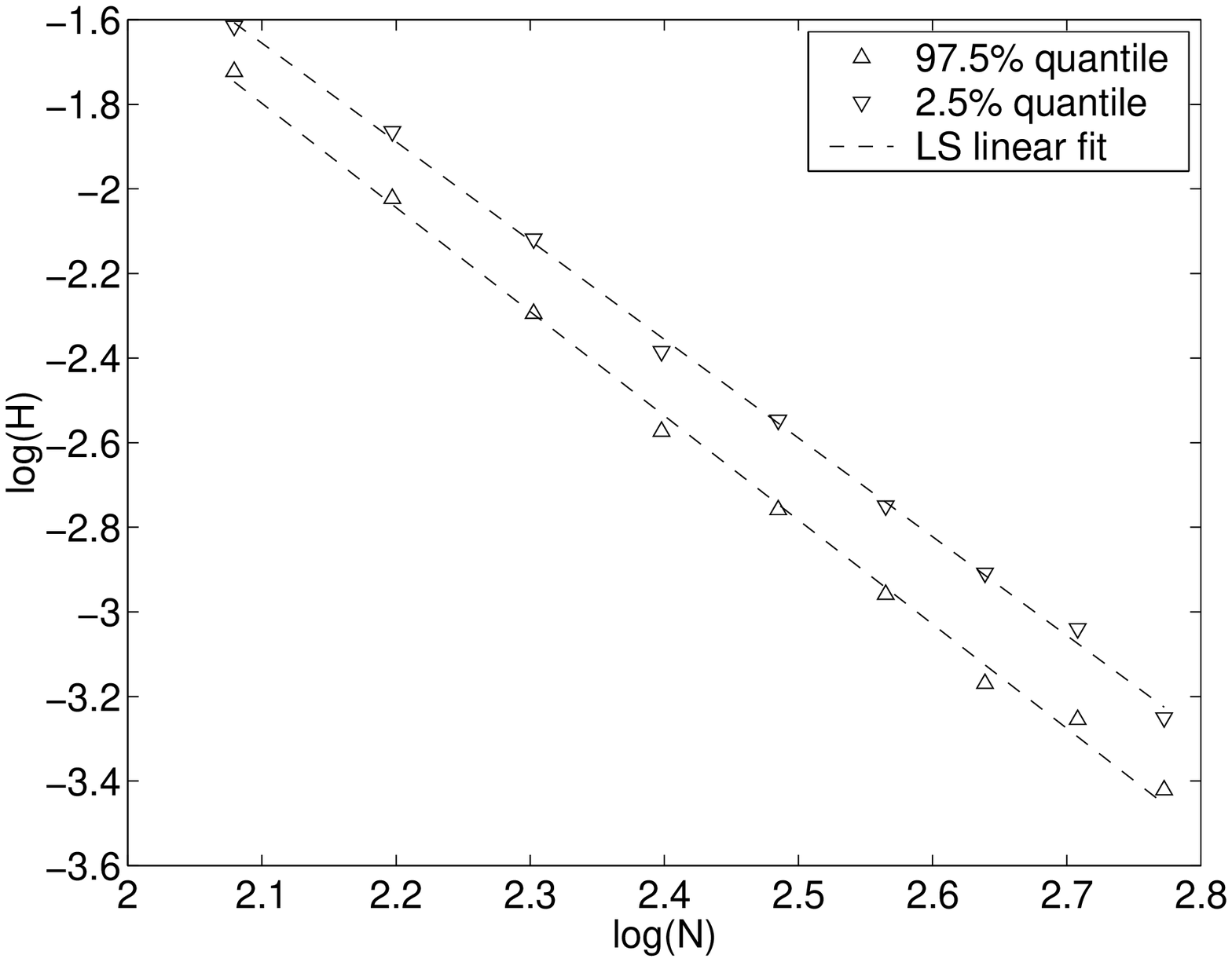}}
\caption{Fitting DFA quantiles ($n>10$): a plot of $\log(\hat{x}_{97.5} - 0.5)$ and 
$\log(0.5 - \hat{x}_{2.5})$ vs. $\log N$, where $N = \log_2 L$.
}

\centerline{\epsfxsize=12cm \epsfbox{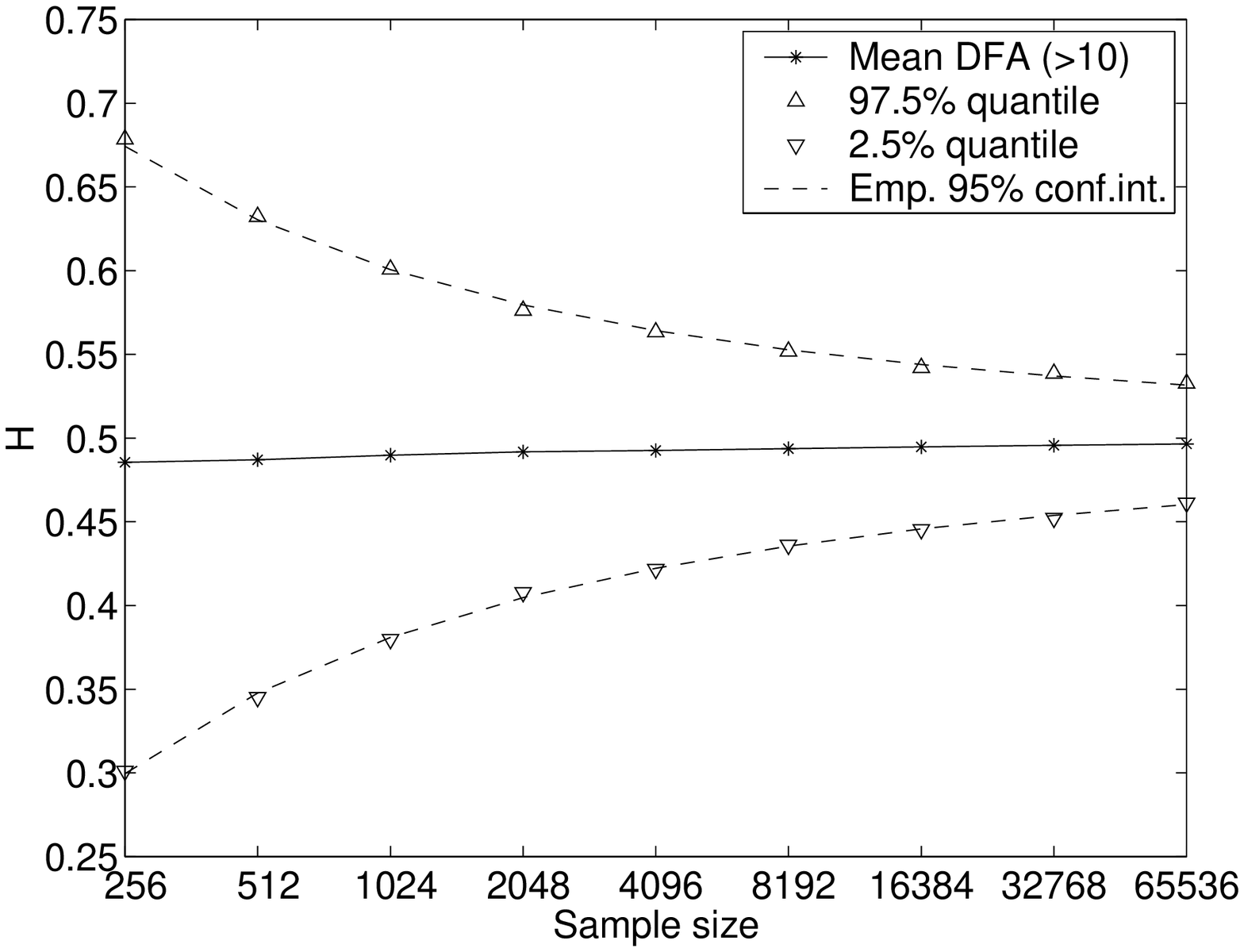}}
\caption{Mean value of the DFA statistics ($n>10$) and 95\% confidence intervals 
for samples of size $L=256, 512, ..., 65536$. 
}
\end{figure}

\begin{figure}[p]
\centerline{\epsfxsize=12cm \epsfbox{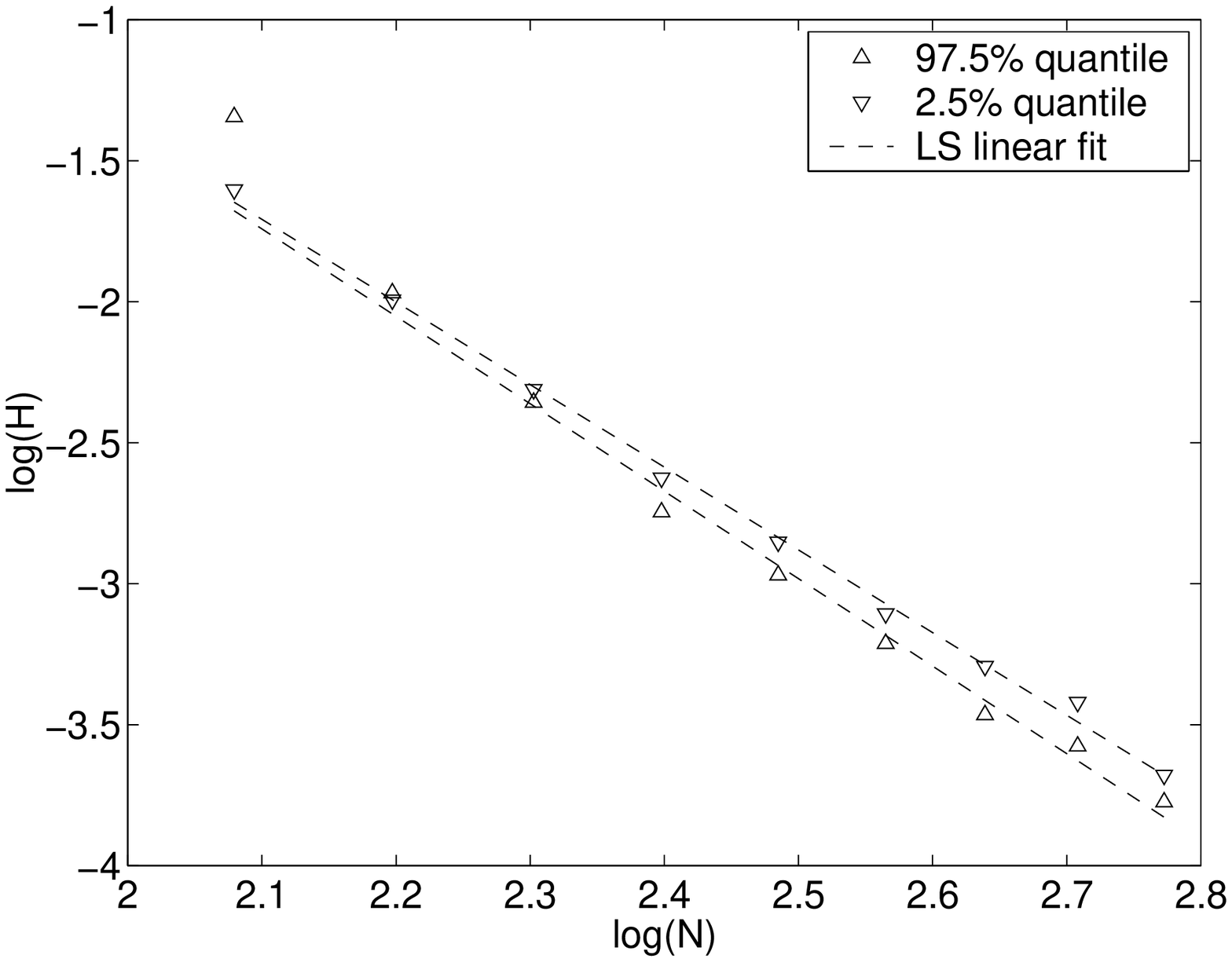}}
\caption{Fitting DFA quantiles ($n>50$): a plot of $\log(\hat{x}_{97.5} - 0.5)$ and 
$\log(0.5 - \hat{x}_{2.5})$ vs. $\log N$, where $N = \log_2 L$.
For the smallest analyzed samples ($L=256$) the 97.5\% quantile introduced large 
estimation errors and was not used in the analysis.
}

\centerline{\epsfxsize=12cm \epsfbox{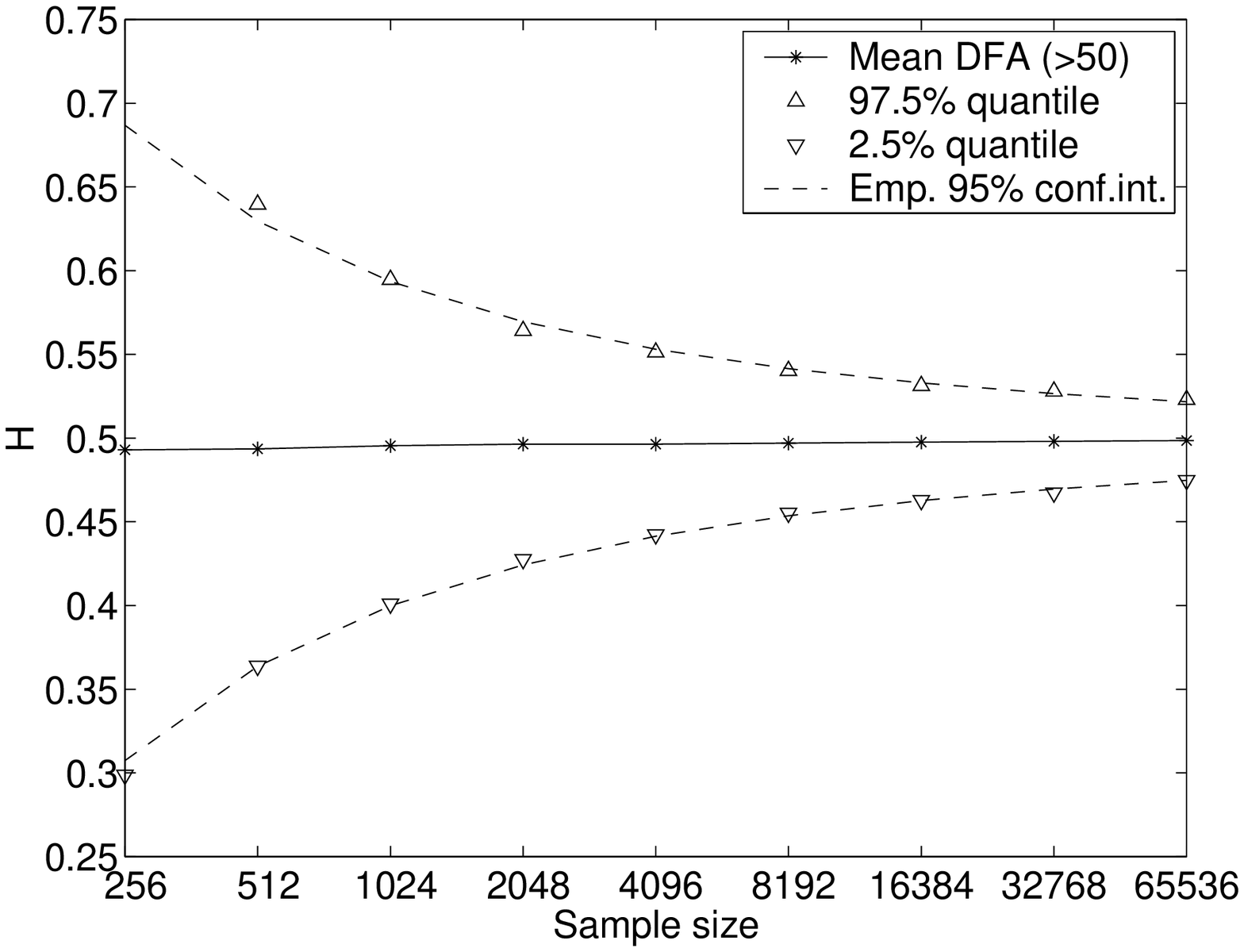}}
\caption{Mean value of the DFA statistics ($n>50$) and 95\% confidence intervals 
for samples of size $L=256, 512, ..., 65536$. 
}
\end{figure}

Finally, results for the GPH statistics are presented in Table 4 and Figs. 10-11. 
The only satisfactory results for the 95\% confidence interval were obtained by plotting
$\log(\hat{x}_{97.5} - 0.5)$ and $\log(0.5 - \hat{x}_{2.5})$ against $N^{2/3}$, where
$N = \log_2 L$, see Fig. 10. The power $\frac23$ was chosen arbitrarily, however, comparably
good results were possible in the range $(0.6,0.7)$. The obtained fit was very good, 
the $R^2$ statistics for the lower quantile was 0.9953 and for the upper 0.9987. 
The complete formulas, also for 90\% and 99\% confidence intervals are given in Table 4. 
In Figure 11 we plotted the mean value of $H$ and the 95\% confidence interval vs. sample 
size. For comparison we also added the theoretical 95\% confidence intervals, see formula 
(\ref{gph_var}). The empirical and theoretical intervals are quite close to each other, 
especially for large samples. For small samples the GPH statistics has a slight negative 
bias and the empirical values are shifted downward. Recall, however, that the theoretical 
results are derived from the limiting behavior and obviously cannot take into account finite
sample properties. 

The small difference between the empirical and theoretical confidence intervals justifies 
our approach and permits us to use in practical applications the empirical confidence 
intervals given in Tables 2-4. 

\begin{table}[htbp]
\caption{Empirical confidence intervals for the GPH statistics and sample length $L=2^N$.
}
\begin{center}
\begin{tabular}{ccc}
\hline
      & \multicolumn{2}{c}{Confidence intervals for $K = [L^{0.5}]$} \\
Level & Lower bound & Upper bound \\
\hline
90\%  & $0.5 - \exp(-0.71 \cdot N^{2/3} + 1.87)$ 
      & $\exp(-0.68 \cdot N^{2/3} + 1.62) + 0.5$  \\
95\%  & $0.5 - \exp(-0.71 \cdot N^{2/3} + 2.04)$ 
      & $\exp(-0.68 \cdot N^{2/3} + 1.78) + 0.5$  \\
99\%  & $0.5 - \exp(-0.73 \cdot N^{2/3} + 2.45)$ 
      & $\exp(-0.65 \cdot N^{2/3} + 1.92) + 0.5$  \\
\hline
\end{tabular}
\end{center}
\end{table}

\begin{figure}[p]
\centerline{\epsfxsize=12cm \epsfbox{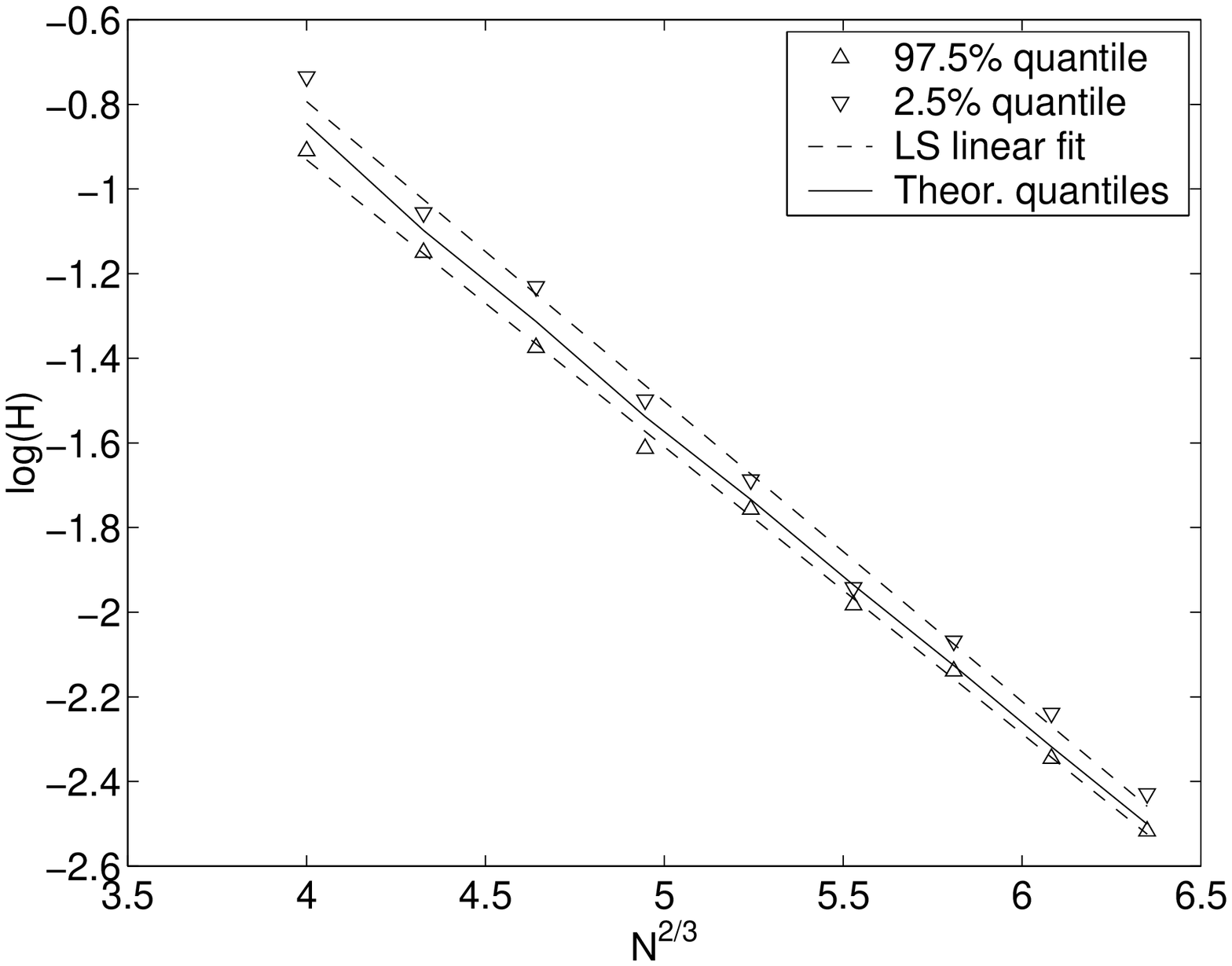}}
\caption{Fitting GPH quantiles: a plot of $\log(\hat{x}_{97.5} - 0.5)$ and 
$\log(0.5 - \hat{x}_{2.5})$ vs. $N^{2/3}$, where $N = \log_2 L$. For comparison 
the theoretical quantiles are also plotted.
}

\centerline{\epsfxsize=12cm \epsfbox{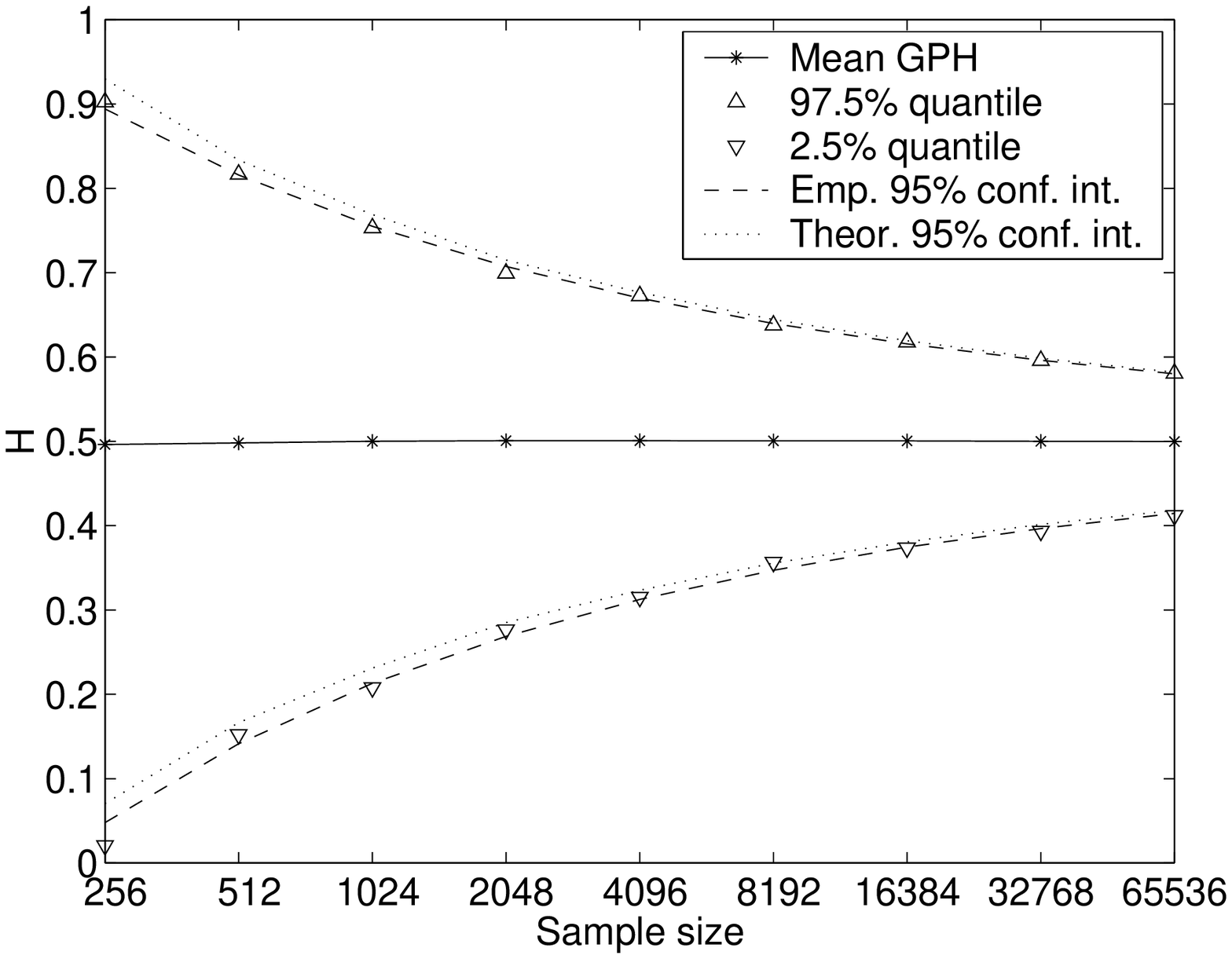}}
\caption{Mean value of the GPH statistics and 95\% confidence intervals for samples 
of size $L=256, 512, ..., 65536$. For comparison the theoretical quantiles are also plotted.
The small difference between the empirical and theoretical confidence intervals justifies 
our approach.
}
\end{figure}

\section{Applications}

Having calculated the empirical confidence intervals we are now ready to test for long
range dependence in financial time series. For the analysis we selected four data sets 
(two stock indices and two power market benchmarks):
\begin{itemize}
\item
2526 daily returns of the Dow Jones Industrial Average (DJIA) index for the period
Jan. 2nd, 1990 -- Dec. 30th, 1999;
\item
1560 daily returns of the WIG20 Warsaw Stock Exchange index (based on 20 blue chip stocks
from the Polish capital market) for the period Jan. 2nd, 1995 -- Mar. 30th, 2001;
\item
728 daily returns of electricity traded in the California Power Exchange (CalPX) spot market
for the period April 1st, 1998 -- March 29th, 2000; 
\item
690 daily returns of firm on-peak power (OTC spot market) in the Entergy region 
(Louisiana, Arkansas, Mississippi and East Texas) for the period January 2nd, 1998 
-- September 25th, 2000.
\end{itemize}

\begin{table}[tbp]
\caption{Estimates of the Hurst exponent $H$ for financial data. $^*$, $^{**}$ and $^{***}$ 
denote significance at the (two-sided) 90\%, 95\% and 99\% level, respectively.}
\begin{center}
\begin{tabular}{llll}
\hline
           & & Method & \\
Data       & R/S-AL & DFA & GPH \\
\hline
 & \multicolumn{3}{c}{\it Stock indices} \\
DJIA returns                    & 0.4585 & 0.4195$^{**}$ & 0.3560 \\
DJIA absolute value of returns  & 0.7838$^{***}$ & 0.9080$^{***}$ & 0.8357$^{***}$ \\
WIG20 returns                   & 0.5030 & 0.4981 & 0.4604 \\
WIG20 absolute value of returns & 0.9103$^{***}$ & 0.9494$^{***}$ & 0.8262$^{***}$ \\
\hline
 & \multicolumn{3}{c}{\it Electricity price benchmarks} \\
CalPX returns           & 0.3473$^*$ & 0.2633$^{***}$ & 0.0667$^{***}$ \\
Entergy returns         & 0.2995$^{**}$ & 0.3651$^{**}$ & 0.0218$^{***}$ \\
\hline
\end{tabular}
\end{center}
\end{table}

The results of the Anis-Lloyd corrected R/S analysis (for $n>50$), the Detrended Fluctuation 
Analysis (for $n>50$ only) and the periodogram Geweke-Porter-Hudak method (for $K=[L^{0.5}]$) 
for these time series are summarized in Table 5. The significance of the results is based
on values obtained from Tables 2-4. For example, the 90\% confidence intervals of the R/S-AL 
statistics for DJIA returns ($N = \log_2 2526 = 11.3026$) were calculated as follows:
\begin{eqnarray*}
\mbox{lower bound:} & 0.5 - \exp(-7.35 \cdot \log(\log N) + 4.06) = 0.4138,\\
\mbox{upper bound:} & \exp(-7.07 \cdot \log(\log N) + 3.75) + 0.5 = 0.5810.
\end{eqnarray*}
Similarly, we obtained confidence intervals for the 95\% and 99\% two-sided levels and 
other methods. All obtained values are presented in Table 6. For comparison we included
the theoretical confidence intervals for the periodogram Geweke-Porter-Hudak method.
The differences between the empirical (obtained from Table 4) and theoretical confidence 
intervals are small and the significance of the results is the same for both sets of values.

\begin{table}[htbp]
\caption{Confidence intervals for 2526 DJIA returns.}
\begin{center}
\begin{tabular}{ccccc}
\hline
 & \multicolumn{4}{c}{Method} \\
Level & R/S-AL & DFA & GPH & GPH (theoretical)\\
\hline
90\% & $(0.4138, 0.5810)$ & $(0.4392, 0.5538)$ & $(0.3184, 0.6645)$ & $(0.3305, 0.6695)$\\
95\% & $(0.3980, 0.5965)$ & $(0.4297, 0.5641)$ & $(0.2847, 0.6931)$ & $(0.2981, 0.7019)$\\
99\% & $(0.3686, 0.6258)$ & $(0.4106, 0.5858)$ & $(0.2067, 0.7583)$ & $(0.2346, 0.7654)$\\
\hline
\end{tabular}
\end{center}
\end{table}

As expected \cite{lux96,ls98,wtt99,g-c00} we found (almost) no evidence for long range 
dependence in the stock indices returns and strong -- i.e. significant at the two-sided 
99\% level for all three methods -- dependence in the stock indices volatility (more precisely:
in absolute value of stock indices returns). 
On the other hand, electricity price returns were found to exhibit a mean-reverting 
mechanism (the Hurst exponent was found significantly smaller than 0.5), which is 
consistent with our earlier findings \cite{wp00}.

\section*{Acknowledgments}

Many thanks to Thomas Lux for stimulating discussions.
Partial financial support for this work was provided by KBN Grant no. PBZ 16/P03/99.

\end{document}